\documentclass[twocolumn,tighten]{aastex62}

\usepackage{graphicx}
\usepackage{ulem}
\usepackage{amsmath}
\usepackage{wrapfig}


\usepackage{amsmath}
\usepackage{graphicx}
\usepackage{dcolumn}
\usepackage{bm}
\usepackage{comment}
\begin{document}



\title{GW190412 as a Third-Generation Black Hole Merger from a Super Star Cluster}

\author[0000-0003-4175-8881]{Carl L.~Rodriguez}
\affil{Harvard Institute for Theory and Computation, 60 Garden St, Cambridge, MA 02138, USA}
\email{carl.rodriguez@cfa.harvard.edu}

\author[0000-0002-4086-3180]{Kyle Kremer}
\affil{ Department of Physics \& Astronomy, Northwestern University, Evanston, IL 60208, USA}
\affil{Center for Interdisciplinary Exploration \& Research in Astrophysics (CIERA), Northwestern University, Evanston, IL 60208, USA}

\author[0000-0002-1655-5604]{Michael Y. Grudi\'c}
\affil{ Department of Physics \& Astronomy, Northwestern University, Evanston, IL 60208, USA}
\affil{Center for Interdisciplinary Exploration \& Research in Astrophysics (CIERA), Northwestern University, Evanston, IL 60208, USA}

\author[0000-0001-7326-1736]{Zachary Hafen}
\affil{ Department of Physics \& Astronomy, Northwestern University, Evanston, IL 60208, USA}
\affil{Center for Interdisciplinary Exploration \& Research in Astrophysics (CIERA), Northwestern University, Evanston, IL 60208, USA}

\author[0000-0002-3680-2684]{Sourav Chatterjee}
\affil{Tata Institute of Fundamental Research, Homi Bhabha Road, Navy Nagar, Colaba, Mumbai 400005, India}

\author[0000-0002-7330-027X]{Giacomo Fragione}
\affil{ Department of Physics \& Astronomy, Northwestern University, Evanston, IL 60208, USA}
\affil{Center for Interdisciplinary Exploration \& Research in Astrophysics (CIERA), Northwestern University, Evanston, IL 60208, USA}

\author[0000-0001-8740-0127]{Astrid Lamberts}
\affil{Universit\'e C\^ote d'Azur, Observatoire de la C\^ote d'Azur, CNRS, Laboratoire Lagrange, Laboratoire ARTEMIS, France}

\author[0000-0001-5285-4735]{Miguel A.~S.~Martinez}

\author[0000-0002-7132-418X]{Frederic A. Rasio}
\affil{ Department of Physics \& Astronomy, Northwestern University, Evanston, IL 60208, USA}
\affil{Center for Interdisciplinary Exploration \& Research in Astrophysics (CIERA), Northwestern University, Evanston, IL 60208, USA}

\author[0000-0002-9660-9085]{Newlin Weatherford}
\affil{ Department of Physics \& Astronomy, Northwestern University, Evanston, IL 60208, USA}
\affil{Center for Interdisciplinary Exploration \& Research in Astrophysics (CIERA), Northwestern University, Evanston, IL 60208, USA}

\author[0000-0001-9582-881X]{Claire S. Ye}
\affil{ Department of Physics \& Astronomy, Northwestern University, Evanston, IL 60208, USA}
\affil{Center for Interdisciplinary Exploration \& Research in Astrophysics (CIERA), Northwestern University, Evanston, IL 60208, USA}

\date{\today}

\begin{abstract}
We explore the possibility that GW190412, a binary black hole merger with a non-equal-mass ratio and significantly spinning primary, was formed through repeated black hole mergers in a dense super star cluster. Using a combination of semi-analytic prescriptions for the remnant spin and recoil kick of black hole mergers, we show that the mass ratio and spin of GW190412 are consistent with a binary black hole whose primary component has undergone two successive mergers from a population of $\sim 10M_{\odot}$ black holes in a high-metallicity environment.  We then explore the production of GW190412-like analogs in the \texttt{CMC Cluster Catalog}, a grid of 148 $N$-body star cluster models, as well as a new model, \texttt{behemoth}, with nearly $10^7$ particles and initial conditions taken from a cosmological MHD simulation of galaxy formation.  We show that{, if the spins of black holes born from stars are small,} the production of binaries with GW190412-like masses and spins is dominated by massive super star clusters with high metallicities and large central escape speeds.  While many are observed in the local universe, our results suggest that a careful treatment of these massive clusters, many of which may have been disrupted before the present day, is necessary to characterize the production of unique gravitational-wave events produced through dynamics.
\end{abstract}



\section{Introduction}
\label{sec:level1}

Since 2015, the LIGO and Virgo gravitational-wave (GW) observatories have reported the detection of 11 binary black hole (BBH) mergers \citep{2015CQGra..32g4001L,2015CQGra..32b4001A,2019PhRvX...9c1040A,2020arXiv200408342T},
with independent analyses \citep{Venumadhav2019,nitz2020} identifying several additional candidates. However, the majority of these previous events have been composed of black holes (BHs) with near-equal component masses\footnote{Though see \cite{Venumadhav2019} for a description of GW170202, a BBH merger candidate with a mass ratio of 2-to-1.}.  This complemented theoretical models of BBH formation through either isolated binary evolution \cite[e.g.,][]{Dominik2012,DeMink2016,2020A&A...636A.104B} or dynamical formation \citep[e.g.,][]{Rodriguez2016} that strongly preferred BBH mergers with mass ratios near unity \citep{Abbott2016x,2019ApJ...882L..24A}. However, the first BBH merger announced from LIGO/Virgo's third observing run ---  GW190412, a $\sim 30M_{\odot}+8M_{\odot}$ binary --- bucks this trend with a mass ratio of nearly 4-to-1 \citep{2020arXiv200408342T}.   This particular configuration allows for the recovery of higher-order modes in the gravitational waveform, which strongly constrains the dimensionless spin magnitude of the primary BH to $0.43\pm^{+0.16}_{-0.26}$, the largest spin of a pre-merger BH measured through GWs.

Although such systems can potentially be formed through isolated binary evolution \cite[e.g.,][]{2020arXiv200409288M,2020arXiv200411866O} or through dynamical exchanges in young, low-mass clusters \cite[e.g.,][]{2020arXiv200409525D}, an obvious way to produce such systems is through hierarchical mergers of BHs in a dense star cluster.  If two low-spinning, ``first-generation'' (1G) BHs were to merge in a cluster with a sufficiently large escape speed, their ``second-generation'' (2G) merger product will remain in the cluster, where it can find another partner and merge again. This scenario has been explored extensively in the literature \citep{Fishbach2017,Gerosa2017} in the context of globular clusters \cite[GCs, ][]{Rodriguez2018,Rodriguez2019}, nuclear star clusters \cite[NSCs,][]{Miller2009,OLeary2009,Antonini2016,Antonini2019}, and the disks of active galactic nuclei \citep{Stone2016,Bartos2016,2020MNRAS.494.1203M,2019PhRvL.123r1101Y,2019ApJ...878...85S,2019ApJ...876..122Y}.  However, the merger of just a single pair of BHs with similar masses and low spins produces BHs with dimensionless spins of $\sim$0.69 \citep{Berti2008,Tichy2008,Kesden2010,Fishbach2017}, a value outside the 90\% posterior probability for the primary spin of GW190412.

In this letter we show that GW190412 is instead consistent with a ``third-generation'' (or 3G) BBH merger whose primary BH was created from two successive BBH mergers.  In Section \ref{sec:clues}, we argue based on previous work in both cluster dynamics and numerical relativity that the most likely dynamical source for GW190412 is a massive, high-metallicity cluster with a large escape speed, and that the retention of 3G BHs in the cluster automatically selects for BHs with spins near the median of the GW190412 primary spin posterior.  In Sections \ref{sec:cmc} and \ref{sec:behemoth}, we use collisional models of star clusters, including a new, massive cluster---\texttt{behemoth}---which produces multiple GW190412-like progenitors that merge in the local universe.  Finally, we conclude by discussing how such massive clusters exist in the local universe, but are strongly dependent on the galactic environments they inhabit.

\section{Clues to the Formation of GW190412}
\label{sec:clues}

  \begin{figure}[]
\centering
\includegraphics[scale=0.9, trim=0.16in 0.21in 0.01in 0.1in, clip=true]{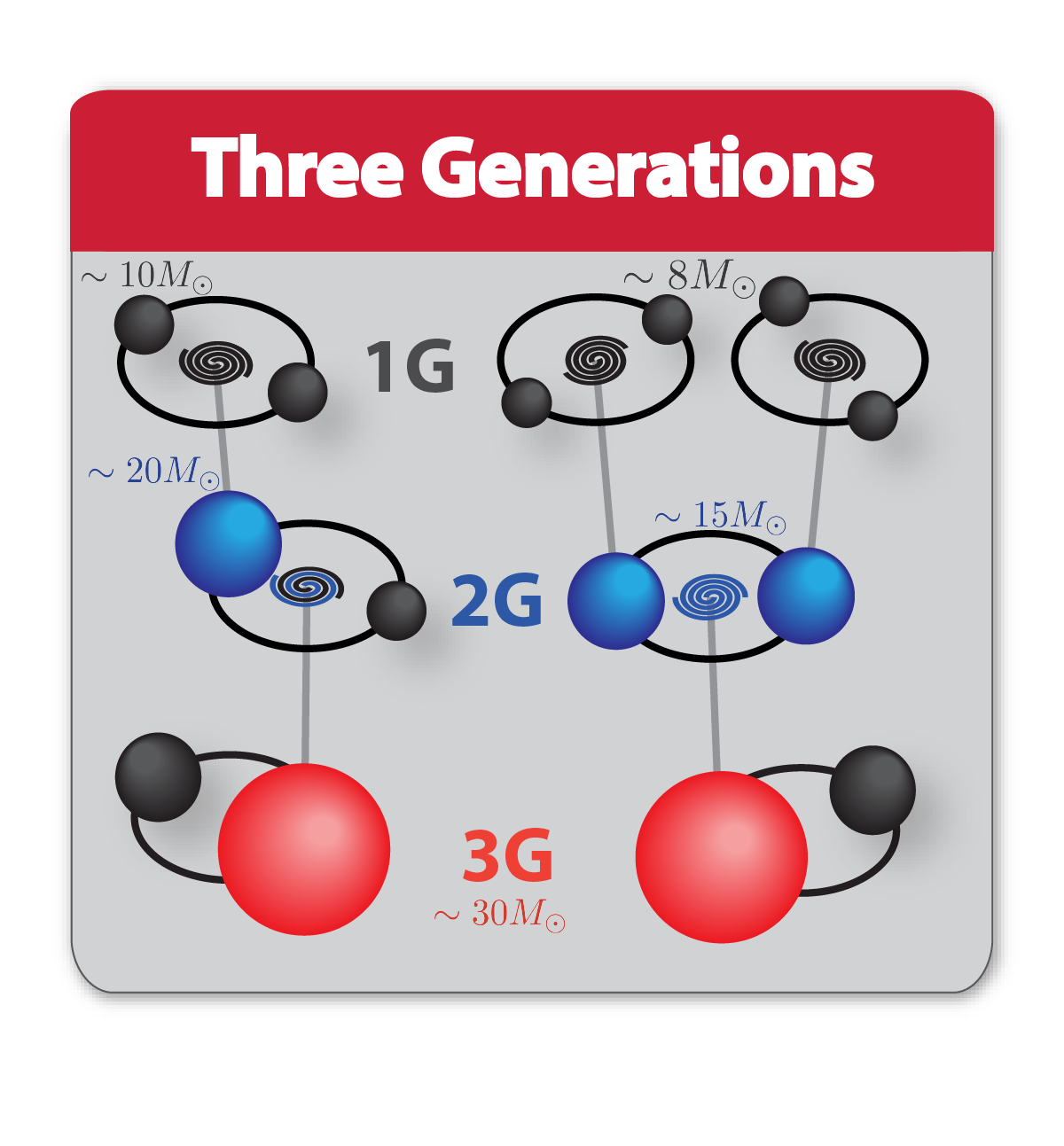}
\caption{A cartoon merger tree of the two possible multi-merger progenitors of GW190412, where the massive primary (in red) is created from the merger of either one or two 2G BHs (in blue)}  
\label{fig:cartoon}
\end{figure}

  \begin{figure}[]
\centering
\includegraphics[scale=0.85, trim=0.in 0.2in 0in 0.4in, clip=true]{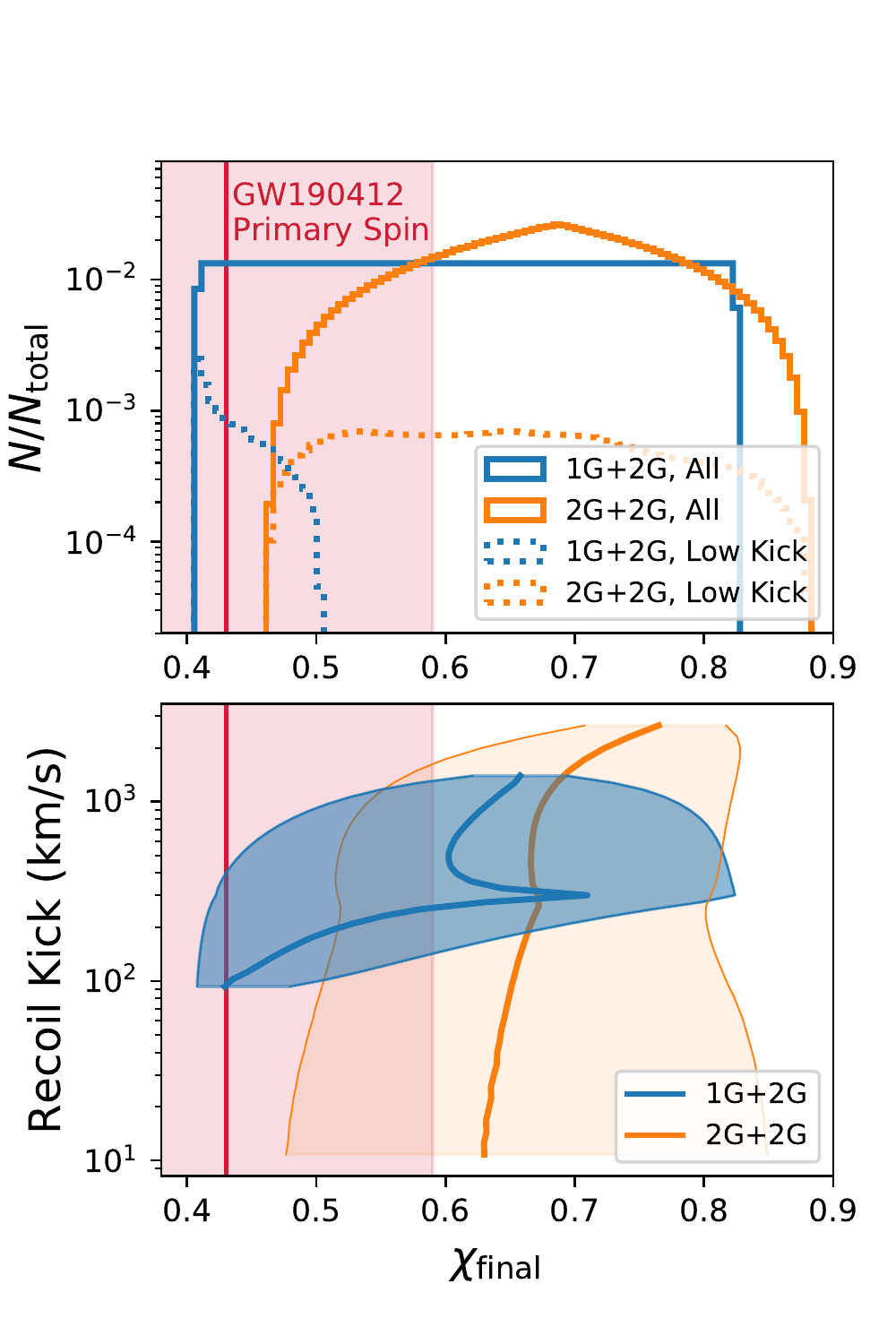}
\caption{Creating GW190412 through the merger of 2G BHs.   The top panel shows the distribution of spins for 3G BHs constructed according to the two schemes in Figure \ref{fig:cartoon}, with the full distribution shown with solid lines and the subset of systems which receive low recoil kicks ($\lesssim 100 \rm{km}/\rm{s}$) shown with dashed lines.  The bottom panel shows the relationship between the 3G BH spins and the recoil kicks they receive at birth, with the solid line and shaded regions showing the median and 90\% regions of allowed final spins (assuming isotropic component spins at merger).  The red line and shaded region indicate the median and 90\% posterior on $\chi_1$ for GW190412.}
\label{fig:kicks}
\end{figure}

Dense star clusters come in a wide range of initial masses, metallicities, and concentrations.  We have previously shown that dense star clusters can naturally form heavy $30M_{\odot}+30M_{\odot}$ BBHs \cite[e.g., GW150914][]{Rodriguez2016b} as well as lower-mass $10M_{\odot}+10M_{\odot}$ binaries \cite[e.g., GW151226,][]{Chatterjee2017}.  The key determining factor between these two mass regimes is the cluster metallicity: low-metallicity systems such as classical globular clusters \cite[GCs, e.g.,][]{Harris1996} are optimal for producing $\sim 30M_{\odot}$ BHs, while higher-metallicity clusters such as open clusters (OCs) and super star clusters \cite[SSCs,][and references therein]{PortegiesZwart2010} preferentially produce lower-mass BHs.  This difference in the BH mass distribution arises from the strength of stellar winds in massive stars; higher metallicity stars experience stronger winds \cite[e.g.,][]{Vink2001} and correspondingly larger mass loss rates, resulting in less massive BH progenitors prior to stellar collapse \cite[e.g.,][]{Belczynski2010,Spera2015a}.  

Of course, a low-metallicity GC containing many $\sim 30 M_{\odot}$ BHs would still contain an even greater population of $\sim 10M_{\odot}$ BHs by virtue of the initial-mass function (IMF), so why do we discount such clusters as a potential source of GW190412-like binaries?  After mass segregation is complete in a star cluster, the most massive BHs are found in the center of the cluster, where they predominantly form binaries with BHs of similar masses \citep{Morscher2015}.  It is these massive BHs that primarily participate in the repeated three-body encounters that form binaries and lead to the ejection of BHs and BBHs from the cluster \citep[e.g.,][]{Kulkarni1993,Sigurdsson1993a}.  Only after the most massive objects have been ejected can the lighter BHs and and neutron stars migrate into the cluster center and participate in such encounters \cite[e.g.][]{Morscher2015,Ye2020}.  Even if---in spite of mass segregation---a 1-to-4 mass ratio binary were to form, approximately 30-50\% of low-velocity encounters between the binary and other massive BHs would result in an ``exchange'' encounter, where the lower-mass BH is replaced by the interloper.   \cite[See][Table 3A]{Sigurdsson1993}.  To eject or merge a BBH from a GC requires 10s to 100s of these low-velocity encounters\footnote{Here ``low-velocity'' refers to those where the interloper is moving below the critical velocity of the system, defined as the velocity where the binding energy of the binary is equal to the kinetic energy of the interloper and the binary center of mass at infinity.  See \cite{Hut1983b}.}, giving many opportunities for an exchange to occur \citep{Rodriguez2016a}.  As long as the BBH components are sourced from a continuous BH mass distribution, mass segregation and three-body dynamics will tend to create equal-mass BBH mergers.

But when a BBH merges inside a cluster, its merger product can be retained by the cluster, where it is nearly twice as massive as the most massive 1G BHs.  These 2G BHs are much more likely than their 1G progenitors to form binaries and merge again within a Hubble time, with a nearly 2-to-1 mass ratio between the 2G and 1G components \citep{Rodriguez2019}.  Furthermore, when two BHs with similar masses merge, the spin distributions of their merger products will be peaked at 0.69 \citep{Berti2007,Fishbach2017}.  This yields an obvious pathway to producing BBHs with a more massive and spinning primary.  However, both a  2-to-1 mass ratio and $\chi_1=0.69$ are excluded at the 90\% level for GW190412.  As we will show, constructing GW190412 from low-spin 1G BHs, with its mass ratio of 0.28 and $\chi_1=0.43$, requires an additional merger.

In Figure \ref{fig:cartoon}, we show two possible pathways to forming $30M_{\odot}$ 3G BHs, where the 3G component is formed through either a 1G+2G merger (with masses $M_1=20M_{\odot}$, $M_2=10M_{\odot}$ and spins $\chi_1=0.69$ and $\chi_2=0$) or a 2G+2G merger (both with masses of $15M_{\odot}$ and spins of $0.69$).  In the top panel of Figure \ref{fig:kicks}, we show the final spin distributions for the resultant 3G BHs, calculated by Monte Carlo sampling over all possible spin orientations of the progenitors using phenomenological fits to numerical and analytic relativity calculations for the final spin and recoil kicks of the BBH merger products \citep{Barausse2009,Campanelli2007,Gonzalez2007,Lousto2008,Lousto2012,Lousto2013}.  Note that these are the same distributions employed in our \texttt{Cluster Monte Carlo} code \citep[][Appendix A]{Rodriguez2018}.

\begin{figure}[]
\centering
\includegraphics[scale=0.85, trim=0.25in 0.in 0in 0.0in, clip=true]{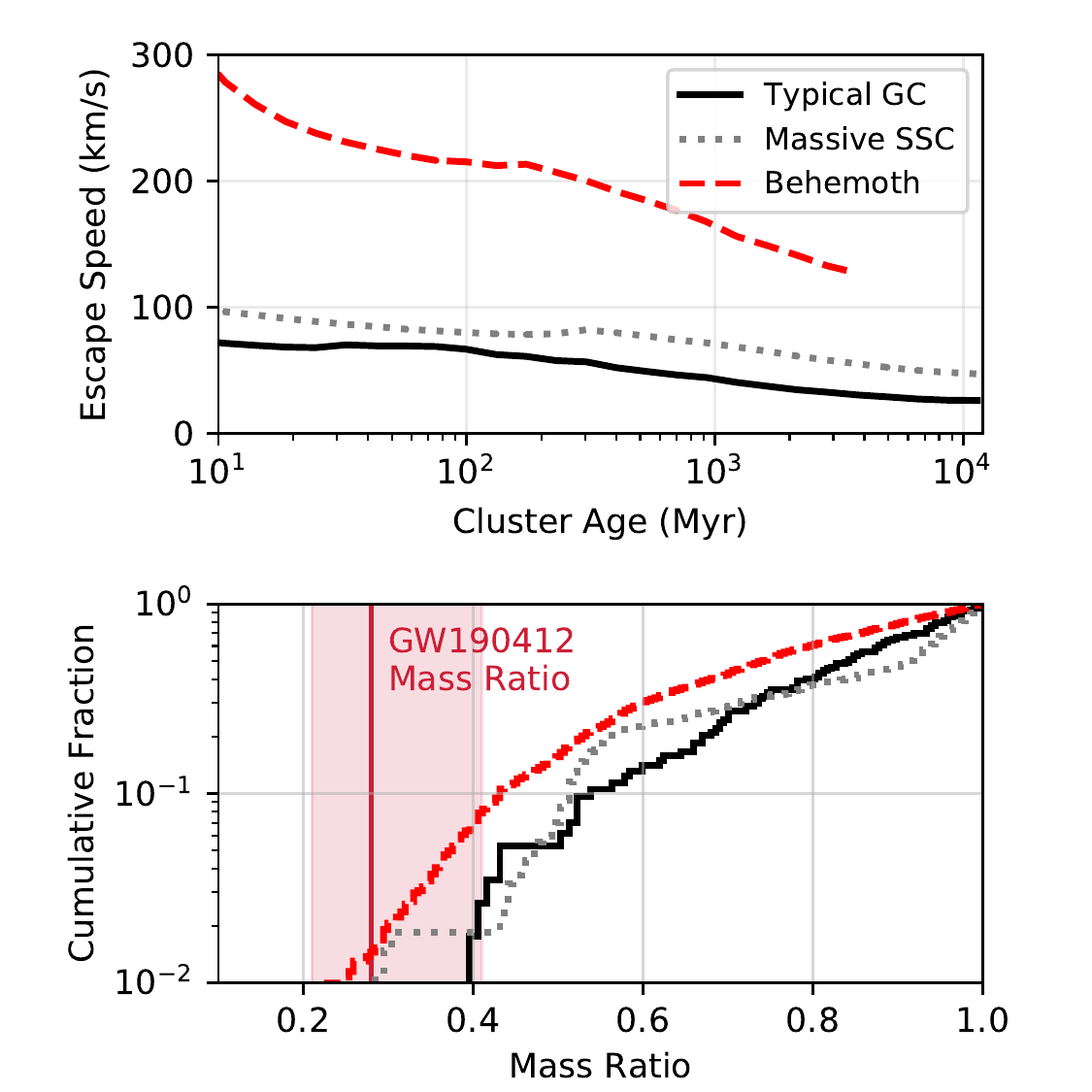}
\caption{\textbf{Top}: the central escape speeds for a typical MW GC (with a metallicity of $Z=0.002$ and a present-day mass of $\sim 2\times10^5M_{\odot}$), a massive SSC that produces a GW190412-like binary ($Z=0.02$ and final mass of  $\sim5\times10^5M_{\odot}$ at the present day), both taken from \cite{Kremer2020}, and \texttt{behemoth}, with metallicity $Z=0.02$ and mass of $\sim 3\times10^6M_{\odot}$ at its time of disruption, which produces several GW190412-like binaries.  \textbf{Bottom}: the cumulative distribution of mass ratios for BBH mergers from the same three clusters, with the median and 90\% mass ratio for GW190412 indicated in red.}
\label{fig:vesc}
\end{figure}

The median final spins for all 1G+2G and 2G+2G mergers ($\chi_f = 0.62$ and $0.68$, respectively) are both beyond the 90\% credible region for the GW190412 primary.  However, if we consider only 3G BHs that receives kicks below 100 $\rm{km}/\rm{s}$, the median remnant spin decreases to $0.43$ and $0.65$ for 1G+2G and 2G+2G mergers, respectively, the former of which agrees perfectly with the median for the spin posterior of the GW190412 primary.    This correlation between BH recoil and remnant spin is shown explicitly in the bottom panel of Figure \ref{fig:kicks}:  100\% of 1G+2G BBH mergers (and $\sim$30\% of 2G+2G mergers) with recoil kicks below $\sim 100 \rm{km}/\rm{s}$ produce BHs with spins matching the LIGO/Virgo posterior.  In other words, while the retention of 3G BHs may be rare, any BHs that are retained will, by selection effects, have spins consistent with the GW190412 primary.  While the agreement between the final spins for 1G+2G mergers would suggest them as the primary source of GW190412-like binaries \citep[especially since they are nearly 13 times more prevalent than 2G+2G mergers,][]{Rodriguez2019}, the minimum recoil speed produced by such binaries is nearly $\sim 90 \rm{km}/\rm{s}$.  This is greater than the escape speed from many nearby GCs and SSCs (although such clusters had significantly larger escape speeds in the past; see the top panel of Figure \ref{fig:vesc}), suggesting that we must expand our search beyond typical Milky Way (MW) clusters.

\section{Star Cluster Models}
\label{sec:cmc}

With a better understanding of the type of cluster most likely to form GW190412, we can search for similar events in high-fidelity models of dense star clusters.  To that end, we use a series of star cluster models created with the \texttt{Cluster Monte Carlo} (CMC) code, a H\'enon-style Monte Carlo code for stellar dynamics \citep{Joshi1999,Joshi2001a,Fregeau2003,Fregeau2007,Umbreit2009,Chatterjee2010,Pattabiraman2013}.  In addition to the orbit-averaged Fokker-Planck diffusion of particles through phase space by two-body encounters \citep{Henon1971,Henon1975} which drives the overall evolution of the cluster, \texttt{CMC} includes all of the necessary physics for treating the formation, dynamics, ejections, and (multiple)  mergers of BBH systems.  This includes probabilistic formation of binaries through three-body BH encounters \citep{Morscher2012}, three- and four-body encounters performed by direct integration \citep{Fregeau2007} and detailed stellar evolution prescriptions for stars and binaries using the Binary Stellar Evolution (BSE) code of \cite{Hurley2000,Hurley2002} with upgraded prescriptions for massive stellar winds and compact-object formation \citep{Rodriguez2016,Rodriguez2018}.  In addition to the mergers of isolated BBHs arising from slow GW emission  (either inside the cluster or after their ejection), \texttt{CMC} also follows the ``prompt'' merger of BBHs that are created by GW emission during two-body BHs encounters in the cluster \cite[following][]{2019arXiv190711231S}, and during three- and four-body strong encounters between binaries \citep{Rodriguez2018}. See \cite{Kremer2020} for a more detailed description.  Note that we assume zero natal spins for 1G BHs \cite[e.g.,][]{fuller2019, fuller2019a}, and do not allow for spin up during BH-star mergers or mass transfer.

We use two sets of initial conditions for our analysis.  The first is the \texttt{CMC Cluster Catalog}, a grid of models developed and presented in \cite{Kremer2020}.  Similar to previous studies, these models covered a wide range of initial particle numbers ($N = 2\times10^5$, $4\times 10^5$, $8\times 10^5$, $1.6\times10^6$, and $3.2\times10^6$), cluster virial radii (0.5, 1, 2, and 4pc), metallicities ($Z = 0.0002$, 0.002, and 0.02), and galactocentric distances (2, 8, and 20 kpc).  Masses of stars are first taken from a \cite{Kroupa2001} initial-mass function, then 5\% of stars are randomly assigned a binary companion with a mass drawn from a flat mass ratio distribution between 0.1 and 1 and semi-major axes drawn from a uniform-in-log distribution \citep{1991A&A...248..485D}.  Each model is integrated for 14 Gyrs or until the cluster is dissolved. 

We also consider one additional model, a massive SSC with a high metallicity named \texttt{behemoth}.  Unlike previous \texttt{CMC} studies based on grids of parameters, this cluster is part of a new survey of realistic cluster initial conditions taken from FIRE-2 MHD cosmological simulations \citep{hopkins2017,hopkins2020_fire_mhdcv}, specifically, the cosmological formation of an $L^{*}$ galaxy \texttt{m12i} \citep{wetzel2016}. This new catalogue will be presented fully in \cite{GrudicInPrep}. \texttt{Behemoth} has an initial $8.6\times10^6$ particles, 10\% of which are binaries, for a total of $9.5\times10^6$ stars with stellar metallicities of $Z = 0.013$.  Following the star-formation and metallicity enrichment history of its host galaxy, \texttt{behemoth} is born at a redshift of 0.78, and unlike previous \texttt{CMC} models, experiences a time-dependent tidal potential based upon the orbit of a tracer particle within the \texttt{m12i} simulation.  

To compute the tidal radius of the cluster, we follow a tracer particle associated with the cluster's formation location within the \texttt{m12i} simulations.  In each snapshot, we compute the local mass, velocity dispersion, and the tidal forces experienced by the particle.  The effective tidal strength is calculated following Appendix D of \cite{pfeffer2018}.  This value is then implemented as the tidal truncation in \texttt{CMC}, where we strip any star whose apocenter moves beyond this boundary. 

The tracer particle of \texttt{behemoth} does not experience dynamical friction, so we must add it in post-processing.   To that end, we again follow \cite{pfeffer2018} and compute the instantaneous timescale for dynamical friction to bring the cluster into its galactic center \citep{Lacey1993} as

\begin{equation}
T_{\rm df} = \frac{\epsilon^{0.78}}{2 B\left(v_c / \sqrt{2}\sigma\right)} \frac{\sqrt{2} \sigma r^2}{G M_c \log \Lambda}~~, 
\label{eqn:tdf}
\end{equation}
  
 \noindent where $r$ is the radius of the orbit in the galaxy, $M_c$ is the cluster mass as a function of time, $v_c$ is the circular velocity of a particle at radius $r$, $B(x) = \rm{erf}(x) - 2x\exp(-x^2)/\sqrt{\pi}$ is the standard dynamical friction velocity expression \cite[e.g.,][]{Binney2011}.  $\epsilon$ is the eccentricity correction from \cite{Lacey1993}, defined as the ratio of the angular momentum of the particle to that of a particle on a circular orbit with the same energy.  Finally, $\log \Lambda$ is the Coulomb logarithm, where $\Lambda = 1+M_c/M_{\rm enc}$, and $M_{\rm enc}$ is mass of the galaxy interior to $r$.  We integrate \texttt{behemoth} until the cumulative number of dynamical friction times is greater than the age of the cluster, i.e.
 
 \begin{equation}
 \int \frac{dt}{T_{\rm df}} > 1~~.
 \label{eqn:intdf}
 \end{equation}
 
 \noindent Once \eqref{eqn:intdf} is satisfied, we assume the cluster has spiraled into its galactic center.  For \texttt{behemoth}, this happens at redshift 0.22, approximately 4.2 Gyrs after the cluster's formation.

\texttt{Behemoth} is the largest cluster model created with \texttt{CMC}, and exhibits some unique properties.   A typical GC produces $\sim 100$ BBHs over a $\sim$12-13 Gyr lifetime, approximately half of which will be ejected from the cluster prior to merger \cite[e.g.,][]{Rodriguez2016,Rodriguez2018}.  But \texttt{behemoth} creates 1300 BBH mergers over its 4.2 Gyr lifetime, 90\% of which merge inside the cluster.  This increase in the fraction of in-cluster mergers is largely due to the large central escape speed of the cluster, initially in excess of 300km$/$s.  As a result, 36\% of BBH mergers from \texttt{behemoth} have at least one component created in a previous merger.  This increase in both in-cluster mergers and higher generation BHs has been noted before in semi-analytic models of massive clusters  such as NSCs \cite[e.g.,][]{Miller2009,Antonini2016}, but this is the first time it has been demonstrated in fully-collisional cluster simulations.

\section{Forming GW190412-like Binaries}
\label{sec:behemoth}

We begin by searching both the grid of models and \texttt{behemoth} for merging binaries whose $m_1$ and $m_2$ lie within the 90\% credible regions for GW190412.  From the \texttt{CMC} grid, we identify 39 such systems which merge within a Hubble time, 36 of which originate from high-metallicity ($Z=0.02$) clusters.  34 of these systems are produced through repeated BH mergers, and as such have spinning primaries.  However, no individual cluster matches \texttt{behemoth}: the massive cluster produces 14 BBH mergers with masses similar to GW190412 over its 4.2 Gyr lifetime, all of which are the result of multiple BH mergers.  The majority of these systems are 1G+2G mergers with mass ratios that sit near the upper 90\% region of the GW190412 mass ratio posterior and primary spins of $\sim 0.69$.  \texttt{Behemoth} produces 3 to 4 times more low-mass-ratio events than a typical GC or massive SSC (Figure \ref{fig:vesc}, bottom panel) with $8\%$ of BBHs having mass ratios consistant with GW190412.

  \begin{figure}[]
\centering
\includegraphics[scale=0.7, trim=0.in 0.in 0in 0.0in, clip=true]{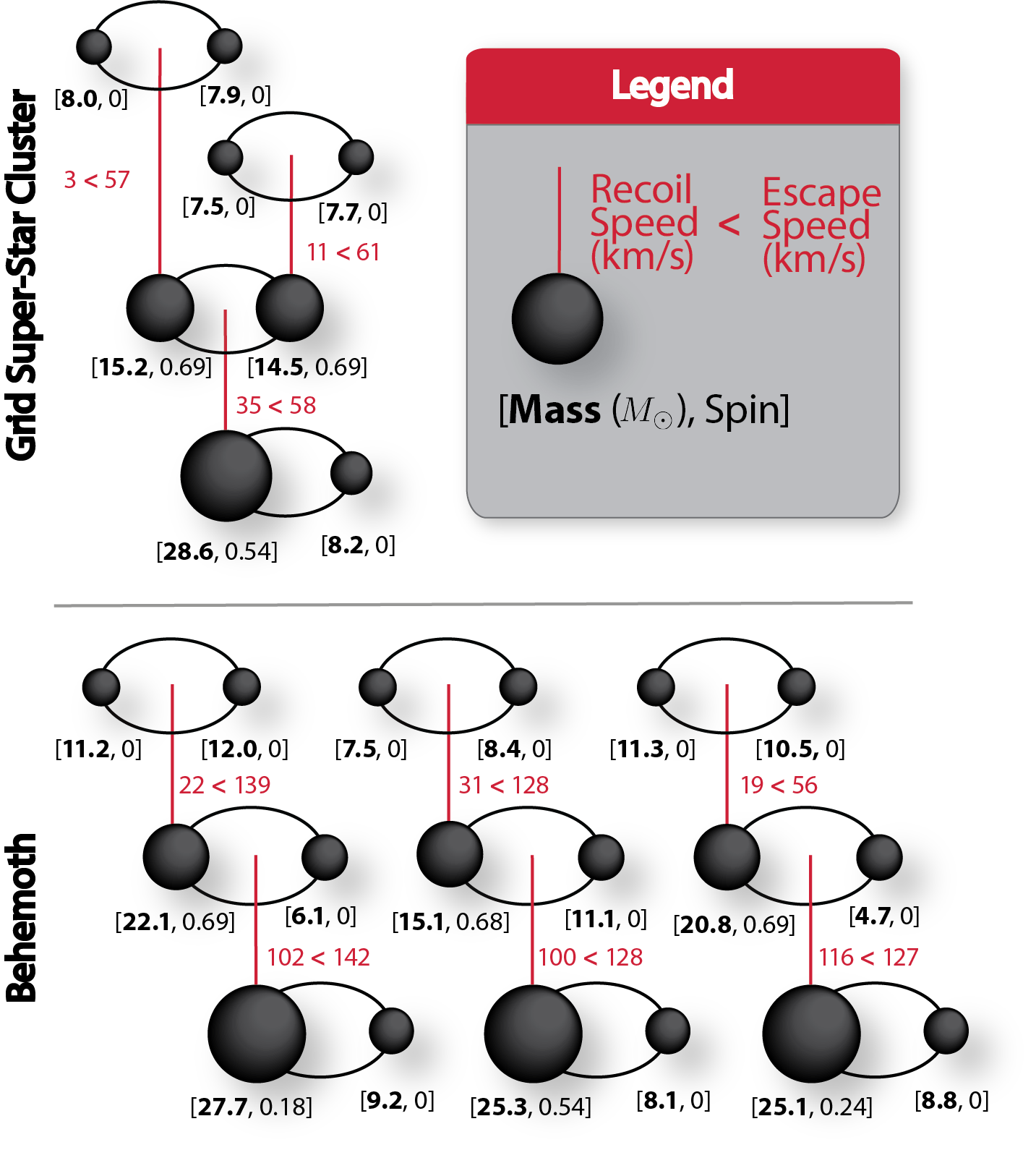}
\caption{BH merger trees that form GW190412-like binaries.  The pairs of numbers under each BH show the mass (in $M_{\odot}$) and spin, while the inequalities show the GW recoil of each merging binary as being less than the local cluster escape speed (in $\rm{km}/\rm{s}$) where the merger occurs.  The top tree shows the one BBH formed through a 2G+2G merger in the grid of cluster models described in \cite{Kremer2020}, while the bottom three trees show the mergers that occur in \texttt{behemoth}, where the large escape speed allows for the retention of 1G+2G merger products.}
\label{fig:trees}
\end{figure}

There are many clusters that can produce BBHs with component masses similar to GW190412.  However, imposing the requirement that the spin magnitude of the primary matches the LIGO/Virgo posterior tells a different story.  Limiting our sample to only those binaries where the primary BH has spin magnitude between 0.17 and 0.59, we find only 4 potential GW190412 progenitors, all of which are 3G+1G BBHs.  Of these, three were created in \texttt{behemoth} through the 1G+2G channel described in Section \ref{sec:clues}, which required an escape speed $\gtrsim 90 \rm{km}/\rm{s}$.  Only one system, where a chance 2G+2G merger experienced a low-recoil kick and was retained by its parent cluster, was produced in the entire cluster grid from \cite{Kremer2020}, despite the grid having more than 10 times the total stellar mass of \texttt{behemoth}.   As expected, this merger occurred in a massive, high-metallicity cluster with $Z=0.02$ and an initial mass of $\sim 10^6M_{\odot}$.  However, the central escape speed of this cluster (shown in Figure \ref{fig:vesc}) falls below the $90 \rm{km}/\rm{s}$ threshold for the 1G+2G pathway within 10s of Myrs of its birth, before the first dynamically-assembled BBHs begin to merge.  It is only in the most massive of high-metallicity clusters where the 1G+2G process for creating GW190412-like binaries  can occur.  We show merger trees for the 4 BBH systems, including their masses, spins, recoil speeds, and local cluster escape speeds, in Figure \ref{fig:trees}.

It is nearly impossible to estimate the volumetric rate of 3G+1G mergers like 
GW190412 using a handful of mergers from two cluster models, and even 
back-of-the-envelope estimates provide limited clarity.  {However, we can 
proceed as follows: for the \texttt{CMC Cluster Catalog}, we compute a 
cosmological merger time for each BBH by adding its \texttt{CMC}-computed merger 
time to a randomly drawn cluster birth redshift based on the metallicitiy of its 
host, using the semi-analytic GC formation models of \cite{El-Badry2018}.  These 
BBH merger times are then combined with the BBHs from \texttt{behemoth}, whose 
merger times are determined by the combination of internal dynamics and the 
cluster's birth redshift in the \texttt{m12i} simulation.\footnote{Though we 
note that the formation redshift of \texttt{behemoth} in the \texttt{m12i} 
simulation, $z=0.78$, is very similar to what \cite{El-Badry2018} would predict 
for a similarly high-metallicitiy cluster ($z=0.81$).}  This population yields 
1757 total BBH mergers at $z\lesssim 0.5$, of which 54 have mass ratios 
$\lesssim 0.41$ (39 of which are produced by \texttt{behemoth}). If we restrict 
ourselves to binaries whose masses individually match the GW190412 posterior 
probability distribution, we are left with 5 BBHs from \texttt{behemoth} and 3 from the \texttt{CMC Cluster Catalog}.  Restricting ourselves further to those whose masses and spins match the GW posterior, only the 3 BBH systems from \texttt{behemoth}, illustrated in Figure \ref{fig:trees}, remain.  If we assume a merger rate of $\lesssim 20 \rm{Gpc}^{-3}\rm{yr}^{-1}$ from clusters at $z < 0.5$ \citep[][Figure 1]{Rodriguez2018b}, this suggests relative merger rates of $\sim 0.6 \rm{Gpc}^{-3}\rm{yr}^{-1}$ for BBHs with mass ratios less than 0.41, $\sim 0.09 \rm{Gpc}^{-3}\rm{yr}^{-1}$ for BBHs with GW190412-like $m_1$ and $m_2$, and $\sim 0.03 \rm{Gpc}^{-3}\rm{yr}^{-1}$ for BBHs with $m_1$, $m_2$, and $\chi_1$ all consistent with GW190412.}


However, {the initial conditions for the \texttt{CMC Cluster Grid} are based upon star clusters in the MW.}  \texttt{Behemoth} is decidedly \emph{not} a typical present-day MW star cluster, globular or otherwise.  {Its  central escape speed is more than $200 \rm{km}/\rm{s}$ after 130 Myr, when it begins to dynamically produce BBH mergers, and remains above $150 \rm{km}/\rm{s}$ after 2 Gyr, when the production of GW190412-like binaries begins.  There are no clusters with central escape speeds in the MW, where even the largest GCs have central escape speeds $\lesssim 125 \rm{km}/\rm{s}$ \citep{2002ApJ...568L..23G}.   However, SSCs with comparable masses and escape velocities are observed in nearby galaxies.  In NGC 3034, the cluster SSC-L has a virial mass of $4\times10^6 M_{\odot}$ \citep{McCrady2007} and an effective radius of $1.45$pc \citep[assuming a distance of 3.6 Mpc,][]{2003ApJ...596..240M}, suggesting a central escape speed of $154 \rm{km}/\rm{s}$.\footnote{Here we calculate the central escape speed as $v^{\rm center}_{\rm{esc}} = \sqrt{2 G M_c / r_{\rm eff}}$, where $M_c$ is the cluster mass and $r_{\rm eff}$ is the effective radius, or the projected half-mass radius.  For a Plummer sphere, this relation is exact; see \cite{Heggie2003}, P.~81.}  NGC 34 and NGC 1316 contain clusters S1 \citep{schweizer:2007.ngc34.sscs} and G114 \citep{bastian:2006.sscs} respectively, which have masses of $\sim 2\times10^7M_{\odot}$, and effective radii $\lesssim 5\rm{ pc}$,\footnote{For S1, the effective radius is inferred to be less than 5 pc based on comparisons to clusters in NGC 3921 \citep{2004AJ....128..202S}, while \cite{bastian:2006.sscs} report a best-fit $r_{\rm eff}$ for G114 of 4.08 pc.} suggesting $v^{\rm center}_{\rm{esc}} \gtrsim 180\rm{km}/\rm{s}.$ for both clusters.   Finally, the cluster W3 in NGC 7252 has a mass of nearly $\sim 10^8M_{\odot}$ and an effective radius of 17.5 pc, with \cite{cabreraziri:2016.ngc7252.w3.vesc} calculating a central escape speed anywhere from 193 to 254 $\rm{km}/\rm{s}$. }

{Each of the aforementioned clusters have sufficiently large $v^{\rm central}_{\rm{esc}}$ to enable the formation of GW190412-like binaries through the channels illustrated in Figure \ref{fig:trees}.  }Furthermore, although no such \texttt{behemoth}-like clusters exist in the MW, this may be because any such clusters would have spiraled into the galactic center many Gyrs ago \cite[e.g.,][]{Gnedin1997}.  The cluster's birth properties and survival time are strongly dependent on the properties of its host galaxy, such as its star-formation rate, tidal field, and metallicity enhancement.  Unlike previous rate estimates of BBH mergers from star clusters \citep[e.g.,][]{PortegiesZwart2000,Rodriguez2016,Askar2016,Hong2018,Choksi2018}, this letter suggests that the production rate of GW190412-like binaries likely depends on clusters which \emph{may no longer exist} in the MW and many other galaxies.   Only a handful of analytic studies \citep[e.g.,][]{Fragione2018} have considered the contribution of such massive clusters that no longer exist, and they find that the contribution from such systems is similar to the contribution from all other clusters combined.

\section{Conclusion}
\label{sec:conclusion}

In this letter, we argued that if the recent LIGO/Virgo BBH, GW190412, were formed through classical three-body encounters in a dense star cluster, then it was likely the product of two successive BH merger events.  Using a combination of analytic prescriptions for BBH recoils and spins, and a series of collisional models of dense star clusters (including a new, massive cluster, \texttt{behemoth}), we show that the primary of GW190412 is typical of BHs formed from the merger of two $\sim 10M_{\odot}$ BHs, whose merger product then merges with \emph{another} $\sim10M_{\odot}$ BH.  While such events are rare, we show that any such mergers retained by the cluster would naturally have masses and spins similar to the components of GW190412.  Although this requires a cluster with an escape speed of at least $90 \rm{km}/\rm{s}$, such clusters are known to exist in nearby galaxies, and may have existed in the MW before being destroyed by dynamical friction.

Although we have focused on dynamical formation in unusually large clusters such as SSCs, there is every reason to expect that similar processes operate in potentially higher numbers in the NSCs that reside in the centers of galaxies.  {For systems without central massive BHs, the higher central escape speeds of NSCs can facilitate many generations of BH mergers \cite[e.g.,][]{Antonini2016}, beyond even the three considered here.  Multiple BH mergers are also easily achievable in environments around supermassive BHs, where the high velocity and escape speeds ($\gtrsim 10^3 \rm{km}/\rm{s}$)  can retain many generations of BHs, whether their progenitor binaries formed through GW-assisted captures \citep{OLeary2009} in BH cusps or through gas-assisted captures in AGN disks \cite[e.g.,][]{2019ApJ...876..122Y,2019PhRvL.123r1101Y}.  We note that the initial escape speed of \texttt{behemoth}, $300 \rm{km}/\rm{s}$, approaches the speed required for a collisional runaway of BHs that may form an intermediate-mass BH \cite{Antonini2019}.  However, clusters with central BHs have distinct dynamical features, including high central velocity dispersions, which may actually inhibit the formation of stellar-mass BBHs through three-body encounters \cite[the rate of which scales as $1/\sigma^9$,][]{Ivanova2005}.}  Unfortunately, star-by-star models of NSC dynamics are still beyond the capabilities of the current generation of direct $N$-body and Hen\'on Monte Carlo codes, making such models difficult to study in detail.  Despite this, significant progress has been made in semi-analytic treatments of the hierarchical BBH merger problem \citep[e.g.,][]{2019PhRvD.100d1301G,2020ApJ...893...35D,2020arXiv200307409A,2020RNAAS...4....2K,2020arXiv200500023K}. 

{Throughout this paper, we have assumed that all BHs born from stellar collapse are born with zero spins.  However, if 1G BHs were born spinning, then it is entirely possible that the only 2G BHs that would be retained are those with spins similar to GW190412.  As this manuscript was being completed, we learned of a similar study by \cite{GerosaInPrep} exploring this scenario.  There, the authors use similar semi-analytic arguments and GW parameter-estimation techniques to study a hierarchical merger scenario for GW190412.  Although they focus on 1G+2G mergers (with varying spins for 1G BHs) as opposed to our 1G+3G scenario, they similarly conclude that producing a GW190412-like BBH through a hierarchical merger scenario would require a cluster with an escape speed $\gtrsim 150 \rm{km}/\rm{s}$.  It is also possible that, if 1G BHs are born with substantial spins, then GW190412 could have originated from a chance low-mass-ratio 1G+1G merger.  We identified 5 such mergers in the \texttt{CMC Cluster Catalog}, suggesting the rate of such events, if 1G BHs were born spinning, is $\sim 0.06 \rm{Gpc}^{-3}\rm{yr}^{-1}$.}

The mass of \texttt{behemoth} approaches that of some central star clusters in dwarf galaxies, suggesting that many interesting BBH mergers may arise from the dividing line between massive GCs, SSCs, and NSCs in galactic centers.  In this case, the tidal forces and galactic environment did not play a significant role in the cluster evolution; the mass, metallicity, and dynamical-friction timescale were the main features contributing to the production of BBH outliers like GW190412.  However, for many clusters, the complicated relationship between clusters and their host galaxies, including a careful treatment of dynamical friction and tidal forces, must be correctly addressed \cite[see e.g.,][]{2019arXiv191205560C}.  Efforts to perform zoom-in collisional cluster simulations from cosmological initial conditions are currently underway \citep{GrudicInPrep,RodriguezInPrep}.

\acknowledgments{CR thanks Hanfei Cui and Peter Phan, participants in the Science Research Mentoring Program (SRMP) at the Center for Astrophysics $\vline$ Harvard \& Smithsonian, as well as the anonymous referee, for useful discussions.
CR was supported by an ITC Postdoctoral Fellowship from Harvard University. MG and GF acknowledges support from a CIERA Fellowship at Northwestern University. SC acknowledges support of the Department of Atomic Energy, Government of India, under project no. 12-R\&D-TFR-5.02-0200. This work was supported in part by NSF Grant AST-1716762 at Northwestern University.  AL acknowledges funding from the Observatoire de la C\^ote d'Azur and the Centre National de la Recherche Scientifique through the Programme National des Hautes Energies and the Programme National de Physique Stellaire.  The computations in this paper were run on the FASRC Cannon cluster supported by the FAS Division of Science Research Computing Group and by the Black Hole Initiative (funded by JTF and GBMF grants), both at Harvard University, and on the San Diego Supercomputing Center Comet cluster under allocation PHY180017 granted by the Extreme Science and Engineering Discovery Environment (XSEDE), supported by the NSF (ACI-1548562).}

\bibliographystyle{aasjournal}


\begin{thebibliography}{}
\expandafter\ifx\csname natexlab\endcsname\relax\def\natexlab#1{#1}\fi

\bibitem[{{Aasi} {et~al.}(2015){Aasi}, {Abbott}, {Abbott}, {Abbott},
  {Abernathy}, {Ackley}, {Adams}, {Adams}, {Addesso}, {Adhikari}, {Adya},
  {Affeldt}, {Aggarwal}, {Aguiar}, {Ain}, {Ajith}, {Alemic}, {Allen},
  {Amariutei}, {Anderson}, {Anderson}, {Arai}, {Araya}, {Arceneaux}, {Areeda},
  {Ashton}, \& {Ast}}]{2015CQGra..32g4001L}
{Aasi}, J., {Abbott}, B.~P., {Abbott}, R., {et~al.} 2015, Classical and Quantum
  Gravity, 32, 074001

\bibitem[{Abbott {et~al.}(2016)Abbott, Abbott, Abbott, Abernathy, Acernese,
  Ackley, Adams, Adams, Addesso, Adhikari, Adya, Affeldt, Agathos, Agatsuma,
  Aggarwal, Aguiar, Aiello, Ain, Ajith, Allen, Allocca, Altin, Anderson,
  Anderson, Arai, Araya, Arceneaux, Areeda, Arnaud, Arun, Ascenzi, Ashton, Ast,
  Aston, Astone, Aufmuth, Aulbert, Babak, Bacon, Bader, Baker, Baldaccini,
  Ballardin, Ballmer, Barayoga, Barclay, Barish, Barker, Barone, Barr,
  Barsotti, Barsuglia, Barta, Bartlett, Bartos, Bassiri, Basti, Batch, Baune,
  Bavigadda, Bazzan, Behnke, Bejger, Belczynski, Bell, Bell, Berger, Bergman,
  Bergmann, Berry, Bersanetti, Bertolini, Betzwieser, Bhagwat, Bhandare,
  Bilenko, Billingsley, Birch, Birney, Biscans, Bisht, Bitossi, Biwer,
  Bizouard, Blackburn, Blair, Blair, Blair, Bloemen, Bock, Bodiya, Boer,
  Bogaert, Bogan, Bohe, Bojtos, Bond, Bondu, Bonnand, Boom, Bork, Boschi, Bose,
  Bouffanais, Bozzi, Bradaschia, Brady, Braginsky, Branchesi, Brau, Briant,
  Brillet, Brinkmann, Brisson, Brockill, Brooks, Brown, Brown, Brown, Buchanan,
  Buikema, Bulik, Bulten, Buonanno, Buskulic, Buy, Byer, Cadonati, Cagnoli,
  Cahillane, Bustillo, Callister, Calloni, Camp, Cannon, Cao, Capano, Capocasa,
  Carbognani, Caride, Diaz, Casentini, Caudill, Cavagli{\`a}, Cavalier,
  Cavalieri, Cella, Cepeda, Baiardi, Cerretani, Cesarini, Chakraborty,
  Chalermsongsak, Chamberlin, Chan, Chao, Charlton, {Chassande-Mottin}, Chen,
  Chen, Cheng, Chincarini, Chiummo, Cho, Cho, Chow, Christensen, Chu, Chua,
  Chung, Ciani, Clara, Clark, Cleva, Coccia, Cohadon, Colla, Collette,
  Cominsky, Jr., Conte, Conti, Cook, Corbitt, Cornish, Corsi, Cortese, Costa,
  Coughlin, Coughlin, Coulon, Countryman, Couvares, Cowan, Coward, Cowart,
  Coyne, Coyne, Craig, Creighton, Cripe, Crowder, Cumming, Cunningham, Cuoco,
  Canton, Danilishin, D'Antonio, Danzmann, Darman, Dattilo, Dave, Daveloza,
  Davier, Davies, Daw, Day, DeBra, Debreczeni, Degallaix, Laurentis,
  Del{\'e}glise, Pozzo, Denker, Dent, Dereli, Dergachev, DeRosa, DeRosa,
  DeSalvo, Dhurandhar, D{\'i}az, Fiore, Giovanni, Lieto, Pace, Palma, Virgilio,
  Dojcinoski, Dolique, Donovan, Dooley, Doravari, Douglas, Downes, Drago,
  Drever, Driggers, Du, Ducrot, Dwyer, Edo, Edwards, Effler, Eggenstein,
  Ehrens, Eichholz, Eikenberry, Engels, Essick, Etzel, Evans, Evans, Everett,
  Factourovich, Fafone, Fair, Fairhurst, Fan, Fang, Farinon, Farr, Farr,
  Favata, Fays, Fehrmann, Fejer, Ferrante, Ferreira, Ferrini, Fidecaro, Fiori,
  Fiorucci, Fisher, Flaminio, Fletcher, Fournier, Franco, Frasca, Frasconi,
  Frei, Freise, Frey, Frey, Fricke, Fritschel, Frolov, Fulda, Fyffe, Gabbard,
  Gair, Gammaitoni, Gaonkar, Garufi, Gatto, Gaur, Gehrels, Gemme, Gendre,
  Genin, Gennai, George, Gergely, Germain, Ghosh, Ghosh, Giaime, Giardina,
  Giazotto, Gill, Glaefke, Goetz, Goetz, Gondan, Gonz{\'a}lez, Castro,
  Gopakumar, Gordon, Gorodetsky, Gossan, Gosselin, Gouaty, Graef, Graff,
  Granata, Grant, Gras, Gray, Greco, Green, Groot, Grote, Grunewald, Guidi,
  Guo, Gupta, Gupta, Gushwa, Gustafson, Gustafson, Hacker, Hall, Hall, Hammond,
  Haney, Hanke, Hanks, Hanna, Hannam, Hanson, Hardwick, Harms, Harry, Harry,
  Hart, Hartman, Haster, Haughian, Heidmann, Heintze, Heitmann, Hello, Hemming,
  Hendry, Heng, Hennig, Heptonstall, Heurs, Hild, Hoak, Hodge, Hofman, Hollitt,
  Holt, Holz, Hopkins, Hosken, Hough, Houston, Howell, Hu, Huang, Huerta, Huet,
  Hughey, Husa, Huttner, {Huynh-Dinh}, Idrisy, Indik, Ingram, Inta, Isa, Isac,
  Isi, Islas, Isogai, Iyer, Izumi, Jacqmin, Jang, Jani, Jaranowski, Jawahar,
  {Jim{\'e}nez-Forteza}, Johnson, Jones, Jones, Jonker, Ju, K, Kalaghatgi,
  Kalogera, Kandhasamy, Kang, Kanner, Karki, Kasprzack, Katsavounidis, Katzman,
  Kaufer, Kaur, Kawabe, Kawazoe, K{\'e}f{\'e}lian, Kehl, Keitel, Kelley, Kells,
  Kennedy, Key, Khalaidovski, Khalili, Khan, Khan, Khan, Khazanov, Kijbunchoo,
  Kim, Kim, Kim, Kim, Kim, Kim, King, King, Kinzel, Kissel, Kleybolte,
  Klimenko, Koehlenbeck, Kokeyama, Koley, Kondrashov, Kontos, Korobko, Korth,
  Kowalska, Kozak, Kringel, Krishnan, Kr{\'o}lak, Krueger, Kuehn, Kumar, Kuo,
  Kutynia, Lackey, Landry, Lange, Lantz, Lasky, Lazzarini, Lazzaro, Leaci,
  Leavey, Lebigot, Lee, Lee, Lee, Lee, Lenon, Leonardi, Leong, Leroy, Letendre,
  Levin, Levine, Li, Libson, Littenberg, Lockerbie, Logue, Lombardi, Lord,
  Lorenzini, Loriette, Lormand, Losurdo, Lough, L{\"u}ck, Lundgren, Luo, Lynch,
  Ma, MacDonald, Machenschalk, MacInnis, Macleod, {Maga{\~n}a-Sandoval}, Magee,
  Mageswaran, Majorana, Maksimovic, Malvezzi, Man, Mandel, Mandic, Mangano,
  Mansell, Manske, Mantovani, Marchesoni, Marion, M{\'a}rka, M{\'a}rka,
  Markosyan, Maros, Martelli, Martellini, Martin, Martin, Martynov, Marx,
  Mason, Masserot, Massinger, {Masso-Reid}, Matichard, Matone, Mavalvala,
  Mazumder, Mazzolo, McCarthy, McClelland, McCormick, McGuire, McIntyre,
  McIver, McManus, McWilliams, Meacher, Meadors, Meidam, Melatos, Mendell,
  {Mendoza-Gandara}, Mercer, Merilh, Merzougui, Meshkov, Messenger, Messick,
  Meyers, Mezzani, Miao, Michel, Middleton, Mikhailov, Milano, Miller,
  Millhouse, Minenkov, Ming, Mirshekari, Mishra, Mitra, Mitrofanov,
  Mitselmakher, Mittleman, Moggi, Mohan, Mohapatra, Montani, Moore, Moore,
  Moraru, Moreno, Morriss, Mossavi, Mours, {Mow-Lowry}, Mueller, Mueller, Muir,
  Mukherjee, Mukherjee, Mukherjee, Mukund, Mullavey, Munch, Murphy, Murray,
  Mytidis, Nardecchia, Naticchioni, Nayak, Necula, Nedkova, Nelemans, Neri,
  Neunzert, Newton, Nguyen, Nielsen, Nissanke, Nitz, Nocera, Nolting,
  Normandin, Nuttall, Oberling, Ochsner, O'Dell, Oelker, Ogin, Oh, Oh, Ohme,
  Oliver, Oppermann, Oram, O'Reilly, O'Shaughnessy, Ott, Ottaway, Ottens,
  Overmier, Owen, Pai, Pai, Palamos, Palashov, Palomba, {Pal-Singh}, Pan,
  Pankow, Pannarale, Pant, Paoletti, Paoli, Papa, Paris, Parker, Pascucci,
  Pasqualetti, Passaquieti, Passuello, Patricelli, Patrick, Pearlstone,
  Pedraza, Pedurand, Pekowsky, Pele, Penn, Perreca, Phelps, Piccinni, Pichot,
  Piergiovanni, Pierro, Pillant, Pinard, Pinto, Pitkin, Poggiani, Popolizio,
  Post, Powell, Prasad, Predoi, Premachandra, Prestegard, Price, Prijatelj,
  Principe, Privitera, Prix, Prodi, Prokhorov, Puncken, Punturo, Puppo,
  P{\"u}rrer, Qi, Qin, Quetschke, Quintero, {Quitzow-James}, Raab, Rabeling,
  Radkins, Raffai, Raja, Rakhmanov, Rapagnani, Raymond, Razzano, Re, Read,
  Reed, Regimbau, Rei, Reid, Reitze, Rew, Reyes, Ricci, Riles, Robertson,
  Robie, Robinet, Rocchi, Rolland, Rollins, Roma, Romano, Romano, Romanov,
  Romie, Rosi{\'n}ska, Rowan, R{\"u}diger, Ruggi, Ryan, Sachdev, Sadecki,
  Sadeghian, Salconi, Saleem, Salemi, Samajdar, Sammut, Sanchez, Sandberg,
  Sandeen, Sanders, Sassolas, Sathyaprakash, Saulson, Sauter, Savage, Sawadsky,
  Schale, Schilling, Schmidt, Schmidt, Schnabel, Schofield, Sch{\"o}nbeck,
  Schreiber, Schuette, Schutz, Scott, Scott, Sellers, Sentenac, Sequino,
  Sergeev, Serna, Setyawati, Sevigny, Shaddock, Shah, Shahriar, Shaltev, Shao,
  Shapiro, Shawhan, Sheperd, Shoemaker, Shoemaker, Siellez, Siemens, Sigg,
  Silva, Simakov, Singer, Singer, Singh, Singh, Singhal, Sintes, Slagmolen,
  Smith, Smith, Smith, Son, Sorazu, Sorrentino, Souradeep, Srivastava, Staley,
  Steinke, Steinlechner, Steinlechner, Steinmeyer, Stephens, Stevenson, Stone,
  Strain, Straniero, Stratta, Strauss, Strigin, Sturani, Stuver, Summerscales,
  Sun, Sutton, Swinkels, Szczepa{\'n}czyk, Tacca, Talukder, Tanner, T{\'a}pai,
  Tarabrin, Taracchini, Taylor, Theeg, Thirugnanasambandam, Thomas, Thomas,
  Thomas, Thorne, Thorne, Thrane, Tiwari, Tiwari, Tokmakov, Tomlinson, Tonelli,
  Torres, Torrie, T{\"o}yr{\"a}, Travasso, Traylor, Trifir{\`o}, Tringali,
  Trozzo, Tse, Turconi, Tuyenbayev, Ugolini, Unnikrishnan, Urban, Usman,
  Vahlbruch, Vajente, Valdes, van Bakel, van Beuzekom, van~den Brand, van~den
  Broeck, {Vander-Hyde}, van~der Schaaf, van Heijningen, van Veggel, Vardaro,
  Vass, Vas{\'u}th, Vaulin, Vecchio, Vedovato, Veitch, Veitch, Venkateswara,
  Verkindt, Vetrano, Vicer{\'e}, Vinciguerra, Vine, Vinet, Vitale, Vo, Vocca,
  Vorvick, Voss, Vousden, Vyatchanin, Wade, Wade, Wade, Walker, Wallace, Walsh,
  Wang, Wang, Wang, Wang, Wang, Ward, Warner, Was, Weaver, Wei, Weinert,
  Weinstein, Weiss, Welborn, Wen, We{\ss}els, Westphal, Wette, Whelan, White,
  Whiting, Williams, Williamson, Willis, Willke, Wimmer, Winkler, Wipf, Wittel,
  Woan, Worden, Wright, Wu, Yablon, Yam, Yamamoto, Yancey, Yap, Yu, Yvert,
  Zadro{\.z}ny, Zangrando, Zanolin, Zendri, Zevin, Zhang, Zhang, Zhang, Zhang,
  Zhao, Zhou, Zhou, Zhu, Zucker, Zuraw, Zweizig, \&
  Collaboration}]{Abbott2016x}
Abbott, B.~P., Abbott, R., Abbott, T.~D., {et~al.} 2016, ApJL, 818, L22

\bibitem[{{Abbott} {et~al.}(2019{\natexlab{a}}){Abbott}, {Abbott}, {Abbott},
  {Abraham}, {Acernese}, {Ackley}, {Adams}, {Adhikari}, {Adya}, {Affeldt},
  {Agathos}, {Agatsuma}, {Aggarwal}, {Aguiar}, {Aiello}, {Ain}, {Ajith},
  {Allen}, {Allocca}, {Aloy}, {Altin}, {Amato}, {Ananyeva}, {Anderson}, {LIGO
  Scientific Collaboration}, \& {Virgo Collaboration}}]{2019ApJ...882L..24A}
{Abbott}, B.~P., {Abbott}, R., {Abbott}, T.~D., {et~al.} 2019{\natexlab{a}},
  \apjl, 882, L24

\bibitem[{{Abbott} {et~al.}(2019{\natexlab{b}}){Abbott}, {Abbott}, {Abbott},
  {Abraham}, {Acernese}, {Ackley}, {Adams}, {Adhikari}, {Adya}, {Affeldt},
  {Agathos}, {Agatsuma}, {Aggarwal}, {Aguiar}, {Aiello}, {Ain}, {Ajith},
  {Allen}, {Allocca}, {Aloy}, {Altin}, {Amato}, {Ananyeva}, {Anderson},
  {Anderson}, {LIGO Scientific Collaboration}, \& {Virgo
  Collaboration}}]{2019PhRvX...9c1040A}
---. 2019{\natexlab{b}}, PRX, 9, 031040

\bibitem[{{Acernese} {et~al.}(2015){Acernese}, {Agathos}, {Agatsuma}, {Aisa},
  {Allemandou}, {Allocca}, {Amarni}, {Astone}, {Balestri}, {Ballardin},
  {Barone}, {Baronick}, {Barsuglia}, {Basti}, {Basti}, {Bauer}, {Bavigadda},
  {Bejger}, {Beker}, {Belczynski}, \& {Bersanetti}}]{2015CQGra..32b4001A}
{Acernese}, F., {Agathos}, M., {Agatsuma}, K., {et~al.} 2015, Classical and
  Quantum Gravity, 32, 024001

\bibitem[{Antonini {et~al.}(2019)Antonini, Gieles, \&
  Gualandris}]{Antonini2019}
Antonini, F., Gieles, M., \& Gualandris, A. 2019, MNRAS, 486, 5008

\bibitem[{Antonini \& Rasio(2016)}]{Antonini2016}
Antonini, F., \& Rasio, F.~A. 2016, ApJ, 831, 187

\bibitem[{{Arca Sedda} {et~al.}(2020){Arca Sedda}, {Mapelli}, {Spera},
  {Benacquista}, \& {Giacobbo}}]{2020arXiv200307409A}
{Arca Sedda}, M., {Mapelli}, M., {Spera}, M., {Benacquista}, M., \& {Giacobbo},
  N. 2020, arXiv e-prints, arXiv:2003.07409

\bibitem[{Askar {et~al.}(2016)Askar, Szkudlarek, {Gondek-Rosi{\'n}ska}, Giersz,
  \& Bulik}]{Askar2016}
Askar, A., Szkudlarek, M., {Gondek-Rosi{\'n}ska}, D., Giersz, M., \& Bulik, T.
  2016, MNRAS: Letters, 464, L36

\bibitem[{Barausse \& Rezzolla(2009)}]{Barausse2009}
Barausse, E., \& Rezzolla, L. 2009, ApJ, 704, L40

\bibitem[{Bartos {et~al.}(2017)Bartos, Kocsis, Haiman, \&
  M{\'a}rka}]{Bartos2016}
Bartos, I., Kocsis, B., Haiman, Z., \& M{\'a}rka, S. 2017, ApJ, 835, 165

\bibitem[{{Bastian} {et~al.}(2006){Bastian}, {Saglia}, {Goudfrooij},
  {Kissler-Patig}, {Maraston}, {Schweizer}, \& {Zoccali}}]{bastian:2006.sscs}
{Bastian}, N., {Saglia}, R.~P., {Goudfrooij}, P., {et~al.} 2006, \aap, 448, 881

\bibitem[{Belczynski {et~al.}(2010)Belczynski, Dominik, Bulik, O'Shaughnessy,
  Fryer, \& Holz}]{Belczynski2010}
Belczynski, K., Dominik, M., Bulik, T., {et~al.} 2010, ApJ, 715, L138

\bibitem[{{Belczynski} {et~al.}(2020){Belczynski}, {Klencki}, {Fields},
  {Olejak}, {Berti}, {Meynet}, {Fryer}, {Holz}, {O'Shaughnessy}, {Brown},
  {Bulik}, {Leung}, {Nomoto}, {Madau}, {Hirschi}, {Kaiser}, {Jones}, {Mondal},
  {Chruslinska}, {Drozda}, {Gerosa}, {Doctor}, {Giersz}, {Ekstrom}, {Georgy},
  {Askar}, {Baibhav}, {Wysocki}, {Natan}, {Farr}, {Wiktorowicz}, {Coleman
  Miller}, {Farr}, \& {Lasota}}]{2020A&A...636A.104B}
{Belczynski}, K., {Klencki}, J., {Fields}, C.~E., {et~al.} 2020, \aap, 636,
  A104

\bibitem[{Berti {et~al.}(2007)Berti, Cardoso, Gonzalez, Sperhake, Hannam, Husa,
  \& Br{\"u}gmann}]{Berti2007}
Berti, E., Cardoso, V., Gonzalez, J.~A., {et~al.} 2007, \prd, 76, 064034

\bibitem[{{Berti} \& {Volonteri}(2008)}]{Berti2008}
{Berti}, E., \& {Volonteri}, M. 2008, \apj, 684, 822

\bibitem[{Binney \& Tremaine(2008)}]{Binney2011}
Binney, J., \& Tremaine, S. 2008, Galactic Dynamics ({Princeton University
  Press}), publication Title: (Second Edition)

\bibitem[{{Cabrera-Ziri} {et~al.}(2016){Cabrera-Ziri}, {Bastian}, {Hilker},
  {Davies}, {Schweizer}, {Kruijssen}, {Mej{\'\i}a-Narv{\'a}ez}, {Niederhofer},
  {Brandt}, {Rejkuba}, {Bruzual}, \&
  {Magris}}]{cabreraziri:2016.ngc7252.w3.vesc}
{Cabrera-Ziri}, I., {Bastian}, N., {Hilker}, M., {et~al.} 2016, \mnras, 457,
  809

\bibitem[{Campanelli {et~al.}(2007)Campanelli, Lousto, Zlochower, \&
  Merritt}]{Campanelli2007}
Campanelli, M., Lousto, C., Zlochower, Y., \& Merritt, D. 2007, ApJ, 659, L5

\bibitem[{Chatterjee {et~al.}(2010)Chatterjee, Fregeau, Umbreit, \&
  Rasio}]{Chatterjee2010}
Chatterjee, S., Fregeau, J.~M., Umbreit, S., \& Rasio, F.~A. 2010, ApJ, 719,
  915

\bibitem[{Chatterjee {et~al.}(2016)Chatterjee, Rodriguez, Kalogera, \&
  Rasio}]{Chatterjee2017}
Chatterjee, S., Rodriguez, C.~L., Kalogera, V., \& Rasio, F.~A. 2016, ApJL,
  836, 1

\bibitem[{Choksi {et~al.}(2018)Choksi, Gnedin, \& Li}]{Choksi2018}
Choksi, N., Gnedin, O.~Y., \& Li, H. 2018, MNRAS, 480, 2343

\bibitem[{{Choksi} \& {Kruijssen}(2019)}]{2019arXiv191205560C}
{Choksi}, N., \& {Kruijssen}, J.~M.~D. 2019, arXiv e-prints, arXiv:1912.05560

\bibitem[{De~Mink \& Mandel(2016)}]{DeMink2016}
De~Mink, S.~E., \& Mandel, I. 2016, MNRAS, 460, 3545

\bibitem[{{Di Carlo} {et~al.}(2020){Di Carlo}, {Mapelli}, {Giacobbo}, {Spera},
  {Bouffanais}, {Rastello}, {Santoliquido}, {Pasquato}, {Ballone}, {Trani},
  {Torniamenti}, \& {Haardt}}]{2020arXiv200409525D}
{Di Carlo}, U.~N., {Mapelli}, M., {Giacobbo}, N., {et~al.} 2020, arXiv
  e-prints, arXiv:2004.09525

\bibitem[{{Doctor} {et~al.}(2020){Doctor}, {Wysocki}, {O'Shaughnessy}, {Holz},
  \& {Farr}}]{2020ApJ...893...35D}
{Doctor}, Z., {Wysocki}, D., {O'Shaughnessy}, R., {Holz}, D.~E., \& {Farr}, B.
  2020, \apj, 893, 35

\bibitem[{Dominik {et~al.}(2012)Dominik, Belczynski, Fryer, Holz, Berti, Bulik,
  Mandel, \& O'Shaughnessy}]{Dominik2012}
Dominik, M., Belczynski, K., Fryer, C., {et~al.} 2012, ApJ, 759, 52

\bibitem[{{Duquennoy} \& {Mayor}(1991)}]{1991A&A...248..485D}
{Duquennoy}, A., \& {Mayor}, M. 1991, \aap, 500, 337

\bibitem[{{El-Badry} {et~al.}(2019){El-Badry}, Quataert, Weisz, Choksi, \&
  {Boylan-Kolchin}}]{El-Badry2018}
{El-Badry}, K., Quataert, E., Weisz, D.~R., Choksi, N., \& {Boylan-Kolchin}, M.
  2019, MNRAS, 482, 4528

\bibitem[{Fishbach {et~al.}(2017)Fishbach, Holz, \& Farr}]{Fishbach2017}
Fishbach, M., Holz, D.~E., \& Farr, B. 2017, ApJ, 840, L24

\bibitem[{{Fragione} \& {Kocsis}(2018)}]{Fragione2018}
{Fragione}, G., \& {Kocsis}, B. 2018, \prl, 121, 161103

\bibitem[{Fregeau {et~al.}(2003)Fregeau, Gurkan, Joshi, \& Rasio}]{Fregeau2003}
Fregeau, J.~M., Gurkan, M.~A., Joshi, K.~J., \& Rasio, F.~A. 2003, arXiv,
  astro-ph, 772

\bibitem[{Fregeau \& Rasio(2007)}]{Fregeau2007}
Fregeau, J.~M., \& Rasio, F.~A. 2007, ApJ, 658, 1047

\bibitem[{{Fuller} \& {Ma}(2019)}]{fuller2019a}
{Fuller}, J., \& {Ma}, L. 2019, \apjl, 881, L1

\bibitem[{{Fuller} {et~al.}(2019){Fuller}, {Piro}, \& {Jermyn}}]{fuller2019}
{Fuller}, J., {Piro}, A.~L., \& {Jermyn}, A.~S. 2019, \mnras, 485, 3661

\bibitem[{Gerosa \& Berti(2017)}]{Gerosa2017}
Gerosa, D., \& Berti, E. 2017, \prd, 95, 124046

\bibitem[{{Gerosa} \& {Berti}(2019)}]{2019PhRvD.100d1301G}
{Gerosa}, D., \& {Berti}, E. 2019, \prd, 100, 041301

\bibitem[{{Gerosa} {et~al.}(2020){Gerosa}, {Vitale}, \& {Berti}}]{GerosaInPrep}
{Gerosa}, D., {Vitale}, S., \& {Berti}, E. 2020,   eprint arXiv:2005.04243
 

\bibitem[{Gnedin \& Ostriker(1997)}]{Gnedin1997}
Gnedin, O.~Y., \& Ostriker, J.~P. 1997, ApJ, 474, 223

\bibitem[{{Gnedin} {et~al.}(2002){Gnedin}, {Zhao}, {Pringle}, {Fall}, {Livio},
  \& {Meylan}}]{2002ApJ...568L..23G}
{Gnedin}, O.~Y., {Zhao}, H., {Pringle}, J.~E., {et~al.} 2002, \apjl, 568, L23

\bibitem[{Gonz{\'a}lez {et~al.}(2007)Gonz{\'a}lez, Sperhake, Br{\"u}gmann,
  Hannam, \& Husa}]{Gonzalez2007}
Gonz{\'a}lez, J.~A., Sperhake, U., Br{\"u}gmann, B., Hannam, M., \& Husa, S.
  2007, Physical Review Letters, 98, 091101

\bibitem[{{Grudic} {et~al.}(2020){Grudic}, {Hafen}, {Rodriguez}, \&
  {Lamberts}}]{GrudicInPrep}
{Grudic}, M., {Hafen}, Z., {Rodriguez}, C., \& {Lamberts}, A. 2020, in prep.

\bibitem[{Harris(1996)}]{Harris1996}
Harris, W.~E. 1996, \aj, 112, 1487

\bibitem[{Heggie \& Hut(2003)}]{Heggie2003}
Heggie, D., \& Hut, P. 2003, The Gravitational Million-Body Problem: A
  Multidisciplinary Approach to Star Cluster Dynamics, by Douglas Heggie and
  Piet Hut. Cambridge University Press, 2003, 372 pp.

\bibitem[{H{\'e}non(1971)}]{Henon1971}
H{\'e}non, M. 1971, Astrophysics and Space Science, 14, 151

\bibitem[{H{\'e}non(1975)}]{Henon1975}
---. 1975, Dynamics of the Solar Systems, doi:1975IAUS...69..133H

\bibitem[{Hong {et~al.}(2018)Hong, Vesperini, Askar, Giersz, Szkudlarek, \&
  Bulik}]{Hong2018}
Hong, J., Vesperini, E., Askar, A., {et~al.} 2018, MNRAS, 480, 5645

\bibitem[{Hopkins(2017)}]{hopkins2017}
Hopkins, P.~F. 2017, arXiv:1712.01294 [astro-ph, physics:physics],
  arXiv:1712.01294

\bibitem[{{Hopkins} {et~al.}(2020){Hopkins}, {Chan}, {Garrison-Kimmel}, {Ji},
  {Su}, {Hummels}, {Kere{\v{s}}}, {Quataert}, \&
  {Faucher-Gigu{\`e}re}}]{hopkins2020_fire_mhdcv}
{Hopkins}, P.~F., {Chan}, T.~K., {Garrison-Kimmel}, S., {et~al.} 2020, \mnras,
  492, 3465

\bibitem[{Hurley {et~al.}(2000)Hurley, Pols, \& Tout}]{Hurley2000}
Hurley, J.~R., Pols, O.~R., \& Tout, C.~A. 2000, MNRAS, 315, 543

\bibitem[{Hurley {et~al.}(2002)Hurley, Tout, \& Pols}]{Hurley2002}
Hurley, J.~R., Tout, C.~A., \& Pols, O.~R. 2002, MNRAS, 329, 897

\bibitem[{Hut \& Bahcall(1983)}]{Hut1983b}
Hut, P., \& Bahcall, J.~N. 1983, ApJ, 268, 319

\bibitem[{Ivanova {et~al.}(2005)Ivanova, Belczynski, Fregeau, \&
  Rasio}]{Ivanova2005}
Ivanova, N., Belczynski, K., Fregeau, J.~M., \& Rasio, F.~A. 2005, MNRAS, 358,
  572

\bibitem[{Joshi {et~al.}(2001)Joshi, Nave, \& Rasio}]{Joshi2001a}
Joshi, K.~J., Nave, C.~P., \& Rasio, F.~A. 2001, ApJ, 550, 691

\bibitem[{Joshi {et~al.}(2000)Joshi, Rasio, Zwart, \&
  Portegies~Zwart}]{Joshi1999}
Joshi, K.~J., Rasio, F.~A., Zwart, S.~P., \& Portegies~Zwart, S. 2000, ApJ,
  540, 969

\bibitem[{{Kesden} {et~al.}(2010){Kesden}, {Sperhake}, \& {Berti}}]{Kesden2010}
{Kesden}, M., {Sperhake}, U., \& {Berti}, E. 2010, \prd, 81, 084054

\bibitem[{{Kimball} {et~al.}(2020{\natexlab{a}}){Kimball}, {Berry}, \&
  {Kalogera}}]{2020RNAAS...4....2K}
{Kimball}, C., {Berry}, C., \& {Kalogera}, V. 2020{\natexlab{a}}, Research
  Notes of the American Astronomical Society, 4, 2

\bibitem[{{Kimball} {et~al.}(2020{\natexlab{b}}){Kimball}, {Talbot}, {Berry},
  {Carney}, {Zevin}, {Thrane}, \& {Kalogera}}]{2020arXiv200500023K}
{Kimball}, C., {Talbot}, C., {Berry}, C. P.~L., {et~al.} 2020{\natexlab{b}},
  arXiv e-prints, arXiv:2005.00023

\bibitem[{{Kremer} {et~al.}(2020){Kremer}, {Ye}, {Rui}, {Weatherford},
  {Chatterjee}, {Fragione}, {Rodriguez}, {Spera}, \& {Rasio}}]{Kremer2020}
{Kremer}, K., {Ye}, C.~S., {Rui}, N.~Z., {et~al.} 2020, \apjs, 247, 48

\bibitem[{{Kroupa}(2001)}]{Kroupa2001}
{Kroupa}, P. 2001, \mnras, 322, 231

\bibitem[{Kulkarni {et~al.}(1993)Kulkarni, Hut, \& McMillan}]{Kulkarni1993}
Kulkarni, S.~R., Hut, P., \& McMillan, S.~J. 1993, Nature, 364, 421

\bibitem[{{Lacey} \& {Cole}(1993)}]{Lacey1993}
{Lacey}, C., \& {Cole}, S. 1993, \mnras, 262, 627

\bibitem[{Lousto \& Zlochower(2008)}]{Lousto2008}
Lousto, C.~O., \& Zlochower, Y. 2008, \prd, 77, 044028

\bibitem[{Lousto \& Zlochower(2013)}]{Lousto2013}
---. 2013, \prd, 87, 084027

\bibitem[{Lousto {et~al.}(2012)Lousto, Zlochower, Dotti, \&
  Volonteri}]{Lousto2012}
Lousto, C.~O., Zlochower, Y., Dotti, M., \& Volonteri, M. 2012, \prd, 85,
  084015

\bibitem[{{Mandel} \& {Fragos}(2020)}]{2020arXiv200409288M}
{Mandel}, I., \& {Fragos}, T. 2020, arXiv e-prints, arXiv:2004.09288

\bibitem[{{McCrady} {et~al.}(2003){McCrady}, {Gilbert}, \&
  {Graham}}]{2003ApJ...596..240M}
{McCrady}, N., {Gilbert}, A.~M., \& {Graham}, J.~R. 2003, \apj, 596, 240

\bibitem[{McCrady \& Graham(2007)}]{McCrady2007}
McCrady, N., \& Graham, J.~R. 2007, ApJ, 663, 844

\bibitem[{{McKernan} {et~al.}(2020){McKernan}, {Ford}, {O'Shaugnessy}, \&
  {Wysocki}}]{2020MNRAS.494.1203M}
{McKernan}, B., {Ford}, K.~E.~S., {O'Shaugnessy}, R., \& {Wysocki}, D. 2020,
  \mnras, 494, 1203

\bibitem[{Miller \& Lauburg(2009)}]{Miller2009}
Miller, M.~C., \& Lauburg, V.~M. 2009, Astrophysical Journal, 692, 917

\bibitem[{Morscher {et~al.}(2015)Morscher, Pattabiraman, Rodriguez, Rasio, \&
  Umbreit}]{Morscher2015}
Morscher, M., Pattabiraman, B., Rodriguez, C., Rasio, F.~A., \& Umbreit, S.
  2015, ApJ, 800, 9

\bibitem[{Morscher {et~al.}(2013)Morscher, Umbreit, Farr, \&
  Rasio}]{Morscher2012}
Morscher, M., Umbreit, S., Farr, W.~M., \& Rasio, F.~A. 2013, ApJ, 763, L15

\bibitem[{Nitz {et~al.}(2020)Nitz, Dent, Davies, Kumar, Capano, Harry, Mozzon,
  Nuttall, Lundgren, \& T{\'a}pai}]{nitz2020}
Nitz, A.~H., Dent, T., Davies, G.~S., {et~al.} 2020, ApJ, 891, 123

\bibitem[{O'Leary {et~al.}(2009)O'Leary, Kocsis, \& Loeb}]{OLeary2009}
O'Leary, R.~M., Kocsis, B., \& Loeb, A. 2009, MNRAS, 395, 2127

\bibitem[{{Olejak} {et~al.}(2020){Olejak}, {Belczynski}, {Holz}, {Lasota},
  {Bulik}, \& {Miller}}]{2020arXiv200411866O}
{Olejak}, A., {Belczynski}, K., {Holz}, D.~E., {et~al.} 2020, arXiv e-prints,
  arXiv:2004.11866

\bibitem[{Pattabiraman {et~al.}(2013)Pattabiraman, Umbreit, Liao, Choudhary,
  Kalogera, Memik, \& Rasio}]{Pattabiraman2013}
Pattabiraman, B., Umbreit, S., Liao, W.-k., {et~al.} 2013, ApJSS, 204, 15

\bibitem[{Pfeffer {et~al.}(2018)Pfeffer, Kruijssen, Crain, \&
  Bastian}]{pfeffer2018}
Pfeffer, J., Kruijssen, J. M.~D., Crain, R.~A., \& Bastian, N. 2018, MNRAS,
  475, 4309

\bibitem[{Portegies~Zwart {et~al.}(2010)Portegies~Zwart, McMillan, \&
  Gieles}]{PortegiesZwart2010}
Portegies~Zwart, S.~F., McMillan, S.~L., \& Gieles, M. 2010, \araa, 48, 431

\bibitem[{Portegies~Zwart \& Mcmillan(2000)}]{PortegiesZwart2000}
Portegies~Zwart, S.~F., \& Mcmillan, S. L.~W. 2000, ApJ, 528, 17

\bibitem[{{Rodriguez} {et~al.}(2020){Rodriguez}, {Grudic}, {Hafen}, \&
  {Lamberts}}]{RodriguezInPrep}
{Rodriguez}, C., {Grudic}, M., {Hafen}, Z., \& {Lamberts}, A. 2020, in prep.

\bibitem[{Rodriguez {et~al.}(2018)Rodriguez, {Amaro-Seoane}, Chatterjee, \&
  Rasio}]{Rodriguez2018}
Rodriguez, C.~L., {Amaro-Seoane}, P., Chatterjee, S., \& Rasio, F.~A. 2018,
  Physical Review Letters, 120, 151101

\bibitem[{Rodriguez {et~al.}(2016{\natexlab{a}})Rodriguez, Chatterjee, \&
  Rasio}]{Rodriguez2016a}
Rodriguez, C.~L., Chatterjee, S., \& Rasio, F.~A. 2016{\natexlab{a}}, \prd, 93,
  084029

\bibitem[{Rodriguez {et~al.}(2016{\natexlab{b}})Rodriguez, Haster, Chatterjee,
  Kalogera, \& Rasio}]{Rodriguez2016b}
Rodriguez, C.~L., Haster, C.-J., Chatterjee, S., Kalogera, V., \& Rasio, F.~A.
  2016{\natexlab{b}}, ApJ, 824, L8

\bibitem[{Rodriguez \& Loeb(2018)}]{Rodriguez2018b}
Rodriguez, C.~L., \& Loeb, A. 2018, ApJ, 866, L5

\bibitem[{Rodriguez {et~al.}(2016{\natexlab{c}})Rodriguez, Morscher, Wang,
  Chatterjee, Rasio, \& Spurzem}]{Rodriguez2016}
Rodriguez, C.~L., Morscher, M., Wang, L., {et~al.} 2016{\natexlab{c}}, MNRAS,
  463, 2109

\bibitem[{{Rodriguez} {et~al.}(2019){Rodriguez}, {Zevin}, {Amaro-Seoane},
  {Chatterjee}, {Kremer}, {Rasio}, \& {Ye}}]{Rodriguez2019}
{Rodriguez}, C.~L., {Zevin}, M., {Amaro-Seoane}, P., {et~al.} 2019, \prd, 100,
  043027

\bibitem[{{Samsing} {et~al.}(2019){Samsing}, {D'Orazio}, {Kremer}, {Rodriguez},
  \& {Askar}}]{2019arXiv190711231S}
{Samsing}, J., {D'Orazio}, D.~J., {Kremer}, K., {Rodriguez}, C.~L., \& {Askar},
  A. 2019, arXiv e-prints, arXiv:1907.11231

\bibitem[{{Schweizer} \& {Seitzer}(2007)}]{schweizer:2007.ngc34.sscs}
{Schweizer}, F., \& {Seitzer}, P. 2007, \aj, 133, 2132

\bibitem[{{Schweizer} {et~al.}(2004){Schweizer}, {Seitzer}, \&
  {Brodie}}]{2004AJ....128..202S}
{Schweizer}, F., {Seitzer}, P., \& {Brodie}, J.~P. 2004, \aj, 128, 202

\bibitem[{{Secunda} {et~al.}(2019){Secunda}, {Bellovary}, {Mac Low}, {Ford},
  {McKernan}, {Leigh}, {Lyra}, \& {S{\'a}ndor}}]{2019ApJ...878...85S}
{Secunda}, A., {Bellovary}, J., {Mac Low}, M.-M., {et~al.} 2019, \apj, 878, 85

\bibitem[{Sigurdsson \& Hernquist(1993)}]{Sigurdsson1993}
Sigurdsson, S., \& Hernquist, L. 1993, Nature, 364, 423

\bibitem[{Sigurdsson \& Phinney(1993)}]{Sigurdsson1993a}
Sigurdsson, S., \& Phinney, E.~S. 1993, ApJ, 415, 631

\bibitem[{Spera {et~al.}(2015)Spera, Mapelli, \& Bressan}]{Spera2015a}
Spera, M., Mapelli, M., \& Bressan, A. 2015, MNRAS, 451, 4086

\bibitem[{Stone {et~al.}(2017)Stone, Metzger, \& Haiman}]{Stone2016}
Stone, N.~C., Metzger, B.~D., \& Haiman, Z. 2017, MNRAS, 464, 946

\bibitem[{{The LIGO Scientific Collaboration} \& {the Virgo
  Collaboration}(2020)}]{2020arXiv200408342T}
{The LIGO Scientific Collaboration}, \& {the Virgo Collaboration}. 2020, arXiv
  e-prints, arXiv:2004.08342

\bibitem[{Tichy \& Marronetti(2008)}]{Tichy2008}
Tichy, W., \& Marronetti, P. 2008, \prd, 78, 081501

\bibitem[{Umbreit {et~al.}(2009)Umbreit, Fregeau, Chatterjee, \&
  Rasio}]{Umbreit2009}
Umbreit, S., Fregeau, J.~M., Chatterjee, S., \& Rasio, F.~A. 2009, arXiv,
  astro-ph.G, 31

\bibitem[{{Venumadhav} {et~al.}(2020){Venumadhav}, {Zackay}, {Roulet}, {Dai},
  \& {Zaldarriaga}}]{Venumadhav2019}
{Venumadhav}, T., {Zackay}, B., {Roulet}, J., {Dai}, L., \& {Zaldarriaga}, M.
  2020, \prd, 101, 083030

\bibitem[{Vink {et~al.}(2001)Vink, {de Koter}, \& Lamers}]{Vink2001}
Vink, J.~S., {de Koter}, A., \& Lamers, H. J. G. L.~M. 2001, \aap, 369, 574

\bibitem[{Wetzel {et~al.}(2016)Wetzel, Hopkins, Kim, {Faucher-Gigu{\`e}re},
  Kere{\v s}, \& Quataert}]{wetzel2016}
Wetzel, A.~R., Hopkins, P.~F., Kim, J.-h., {et~al.} 2016, ApJ, 827, L23

\bibitem[{{Yang} {et~al.}(2019{\natexlab{a}}){Yang}, {Bartos}, {Haiman},
  {Kocsis}, {M{\'a}rka}, {Stone}, \& {M{\'a}rka}}]{2019ApJ...876..122Y}
{Yang}, Y., {Bartos}, I., {Haiman}, Z., {et~al.} 2019{\natexlab{a}}, \apj, 876,
  122

\bibitem[{{Yang} {et~al.}(2019{\natexlab{b}}){Yang}, {Bartos}, {Gayathri},
  {Ford}, {Haiman}, {Klimenko}, {Kocsis}, {M{\'a}rka}, {M{\'a}rka}, {McKernan},
  \& {O'Shaughnessy}}]{2019PhRvL.123r1101Y}
{Yang}, Y., {Bartos}, I., {Gayathri}, V., {et~al.} 2019{\natexlab{b}}, \prl,
  123, 181101

\bibitem[{{Ye} {et~al.}(2020){Ye}, {Fong}, {Kremer}, {Rodriguez}, {Chatterjee},
  {Fragione}, \& {Rasio}}]{Ye2020}
{Ye}, C.~S., {Fong}, W.-f., {Kremer}, K., {et~al.} 2020, \apjl, 888, L10

\end{thebibliography}

\end{document}